%% file: samplepaper.tex
\documentclass[runningheads]{llncs}
\usepackage{graphicx}
\usepackage{booktabs}
\usepackage{multirow}
\usepackage{amsmath}
\usepackage{amsfonts}

\begin{document}

\title{An Experimental Study on Pretraining Transformers from Scratch for IR}

\titlerunning{Pretraining Transformers from Scratch}
%
%

\author{Carlos Lassance \and
Hervé Dejean  \and
Stéphane Clinchant}


%
\authorrunning{C. Lassance et al.}

%
\institute{Naver Labs Europe, Meylan, France \\ \email{\{carlos.lassance,herve.dejean,stephane.clinchant\}@naverlabs.com}}

%
%
\maketitle              
\begin{abstract}
Finetuning Pretrained Language Models (PLM) for IR has been {de facto} the standard practice since their breakthrough effectiveness few years ago.
But, is this approach well understood? 
In this paper, we study the impact of the pretraining collection on the final IR effectiveness. In particular, we challenge the current hypothesis that PLM shall be trained on a large enough generic collection and we show that pretraining from scratch on the collection of interest is surprisingly competitive with the current approach.
 We benchmark first-stage ranking rankers and cross-encoders for reranking on the task of general passage retrieval on MSMARCO, Mr-Tydi for Arabic, Japanese and Russian, and TripClick for specific domain.
Contrary to popular belief, we show that, for finetuning first-stage rankers, models pretrained solely on their collection have equivalent or better effectiveness compared to more general models. 
However, there is a slight effectiveness drop for rerankers pretrained only on the target collection. 
Overall, our study sheds a new light on the role of the pretraining collection and should make our community ponder on  building specialized models by
pretraining from scratch. Last but not least, doing so could enable better control of efficiency, data bias and replicability, which are key research questions for the IR community.


\end{abstract}

\section{Introduction}
Transformers models are the main breakthrough in artificial intelligence over the past five years. Pretraining transformers models with Masked Language Modeling (MLM), a form of self-supervision, as proposed in the seminal BERT model~\cite{bert} led to  major improvement in natural language processing, computer vision and other domains. Pretraining and self-supervision have then paved the way to a race on bigger foundation models. Information Retrieval (IR) followed the same trajectory, where Pretrained Language Models (PLM)
have largely outperform previous neural model~\cite{splade_sigir22,gao-callan-2022-unsupervised,lin2020IRBertReview,izacard2021contriever} but also  traditional bag-of-words approaches such as BM25~\cite{robertson1996okapi}.
These advances were all made possible by a combination of large datasets, PLMs such as BERT, but also priors coming from traditional IR methods such as BM25.

However, these PLMs do not perform well out-of-the box for Information Retrieval (on the contrary to NLP \footnote{For instance, freezing the BERT encoding and learning an additional linear layer is sufficient to obtain good performance in NLP \cite{bert}, while such approach is not as effective in IR.}) as
they require a significant fine-tuning procedure.
In IR, there are two types of PLM: the cross-encoders for  reranking a set of top k documents and the dual encoders to deal with an efficient first-stage retrieval~\cite{lin2020IRBertReview}. 
However, the standard pretraining MLM task may not be the best task for IR as argued in~\cite{gao-callan-2021-condenser}. More-so, a growing tendency is to introduce a ``middle-training'' step~\cite{gao-callan-2021-condenser} to bridge this gap and adapt the PLM not only to the retrieval domain, but also to the way the sentences will be encoded. For example~\cite{gao-callan-2022-unsupervised,gao-callan-2021-condenser,ma2020prop} adapt the Masked Language Modeling (MLM) loss, so that the model learns to condense information on the CLS token, which will be used as the de-facto sentence embedding during fine-tuning. Similarly, several middle training tasks have been proposed to better fit the IR tasks or by using weak supervision.

On the one hand, it seems that there is a widespread belief that the downstream effectiveness is essentially due to pretraining on a \textit{large} external collection. For instance, the foundation models report \cite{foundations_report} state that ``AI is undergoing a paradigm shift with the rise of models (e.g., BERT, DALL-E, GPT-3) that are trained on broad data at scale and are adaptable to a wide range of downstream tasks.'' On the other hand, the middle training process seems to contradict the former. This is why, in this paper, we aim to investigate the following questions: 
do we actually need large scale pretraining for Neural IR?
How much knowledge is actually encoded from the large pretraining collection? Besides, what is known about pretraining in IR is largely limited by the MSMARCO setting and therefore 
 a related question is how one shall address  pretraining language models for new languages or new domains when it comes to IR. 

In this paper, we aim to verify if these preconceived notions are needed, or if we could just have combined the pretraining and middle training steps to generate PLMs that are already adapted to the problem at hand with a smaller cost than doing both separately. Overall, this paper makes the following contributions:

\begin{enumerate}
    \item We study pretrained transformers from scratch on IR collections;
    \item We show that first-stage rankers, pretrained on MSMARCO, are as effective or even better in-domain (MSMARCO), while out-of-domain those models generalize as well (sparse retrieval) or worse (dense retrieval);
    \item We evaluate cross encoders that are pretrained from scratch and verify that they actually benefit from external pretraining;
    \item We show that first-stage retrievers, trained from scratch on the target collection, are competitive or outperform domain specific models (e.g. SciBERT on TripClick) and multilingual models (e.g. MContriever on Mr. TiDy);
    \item Variants of Transformers architectures, such as DeBERTa, alleged to be better in NLP benchmarks do not bring benefits to IR, even when trained from scratch.
\end{enumerate}

\section{Related Work}

\paragraph{PLMs in IR:} today, the standard practice of many IR researchers is to simply download an existing pretrained model in order to finetune it on their retrieval task. After their success on reranking, PLMs have been adopted for first-stage ranking with a bi-encoder network  to tackle efficiency requirements~\cite{karpukhin-etal-2020-dense,lin2020IRBertReview,sentence_bert} or with late interaction~\cite{colbert}.
Several training strategies have  been proposed to improve the effectiveness of bi-encoders, such as distillation~\cite{hofstaetter2020_crossarchitecture_kd,lin-etal-2021-batch,Hofstaetter2021_tasb_dense_retrieval,santhanam2021colbertv2} and hard negative mining~\cite{xiong2021approximate,rocketqa_v1,ren-etal-2021-rocketqav2}.
Parallel to these developments, another research direction aimed at learning \textit{sparse} representations from PLM to behave as lexical matching. COIL~\cite{gao-etal-2021-coil} (later improved in uniCOIL~\cite{DBLP:journals/corr/abs-2106-14807}) learns term-level dense representations to perform contextualized lexical match.
SPLADE~\cite{splade_sigir22,efficiency_splade_22} directly learns high-dimensional sparse representation thanks to the MLM head of the PLM and the help of sparse regularization. Most notably, SPLADE achieved state-of-the art effectiveness on the zero-shot benchmark BEIR~\cite{beir}, being later surpassed by other methods with much more compute~\cite{muennighoff2022sgpt}.

\paragraph{Rise of middle training: } several works recently proposed to perform an additional step of pretraining, before the final finetuning stage, a procedure that we will call  here \textit{middle training}. The rationale is that the PLM weights or its CLS pooling mechanism are not well-suited for retrieval or similarity tasks often used in IR. Two main ideas emerge from this literature: i) using a contrastive loss on different document spans, and ii) using an information bottleneck to better pre-condition the network to rely on its CLS representation to perform predictions~\cite{kim-etal-2021-self}. 
In~\cite{Chang2020Pre-training_ICLR}, the paper compares the Inverse Cloze Task,  Wiki Link Prediction and Body First Selection. Their result show that a combination of all these tasks was beneficial compared to MLM pretraining only. In~\cite{Ma2021Pretraining_hyperlink}, hyperlinks are used for pretraining. Another pretraining relies on web page structure and their DOM in the WebFormer model~\cite{webformer_sigir22}.
Furthermore, Contriever~\cite{izacard2021contriever} relies on
contrastive loss from different text spans, similarly to Co-Condenser 
\cite{gao-callan-2022-unsupervised}. Co-Condenser extends the Condenser 
\cite{gao-callan-2021-condenser} idea which focuses on middle training the CLS token. Very recently, Retro-MAE
\cite{retromae_arxiv22} revisits the same idea, by masking twice an input passage so that the first masking produce a CLS representation reused for decoding the second masking of the passage. Pretraining for sparse models has been recently investigated in~\cite{efficiency_splade_22} to better condition the network with SPLADE finetuning. The idea is to reuse the FLOPS~\cite{paria2020minimizing} regularization used during finetuning within the MLM middle training.
In~\cite{losses_pretraining_2020}, 14 different pretraining tasks are compared when training BERT models, including predicting the \textit{tfidf} scores of a document, which was shown to be beneficial. They then evaluate on several NLP tasks, including sentence similarity. In addition,~\cite{ma2020prop,ma2020bprop} propose pretraining with representative word predictions: for each document a set of important word is defined by several heuristics and the model is pretrained to predict that set of words.

While pretraining for IR  seems very trendy, the idea of pretraining representation for IR tasks can actually be traced before the advent of PLM (or Foundation Models). For instance, more than ten years ago, the supervised semantic indexing model~\cite{ssi_2009},  used hyperlinks anchor to build triplets in a contrastive task, which can be viewed as an ancestor to the pretrainings tasks on Wikipedia. Similarly, weak supervision coming from BM25 was used to pretrain neural IR models~\cite{nrm_weak_supervision_2017} before the use of PLM. 

\paragraph{Foundational models and architectures: } a  loosely related line of work is the scaling laws literature for large pretrained language models~\cite{Kaplan2020ScalingLF}. The scaling law aims to understand how the architecture and model size influence the perplexity and accuracy of the language model. In~\cite{Tay2022ScaleEI}, Tay et al. showed that perplexity was a poor predictive measure of downstream effectiveness and propose to favor depth rather than just width (i.e. hidden size) in a pattern they named \textit{deep-narrow} architectures. While this question could be interesting to our work, most of the literature is focusing on the large data and model regime, while in IR we look to the other side of the spectrum with small or moderate size collections/models for efficiency purposes.

Finally, a very interesting work by Tay et al.~\cite{tay-etal-2021-pretrained}
argued that pre-training and architectural advances have been conflated. In addition, they show that convolutional models are in fact competitive with transformers when they are pre-trained on the same collection and for tasks which do not require cross attention between two sentences (e.g a bi-encoder network). Finally, they argued that the current approach is misguided and that both architecture and pretraining should be considered independently. Finally, \cite{bert_nmt_2019} studies the impact of the pretraining collection for machine translation and in \cite{el2021large} for image related tasks with large models. Our work is related to those, as we study the impact of the pretraining collection for the final effectiveness of an IR system.

\section{Pretraining from Scratch}

All in all, the role of finetuning representation or performing a middle training seems important for IR systems.
The PLM representations seem critical but for good effectiveness, the impact of the pretraining collection on the final effectiveness is unclear since these representations are then finetuned by diverse means. Is the effectiveness heavily influenced by the pretraining collection and its co-occurrence statistics? To investigate this question, we experiment with several PLM models \textbf{trained from scratch} on the target collection. This way, we will be able to measure the effectiveness gains obtained by pretraining on a larger external collection and answer the following questions: is there an advantage in pretraining directly and only on the target collection? When and what are the advantages of pretraining on a different larger collection?

On the one hand, pretraining from scratch on the target collection could have the advantage of better modelling the target collection by having more informative co-occurrence statistics between tokens. Moreover, 'smaller' sized models may be able to reach the same level of effectiveness, i.e. one can also include efficiency requirements when training these models from scratch. On the other hand, pretraining on a larger collection may lead to more robust/generic as the model has seen more domains, different token usages and could have 'more' knowledge.

\begin{figure}
    \centering
    \includegraphics[width=\textwidth]{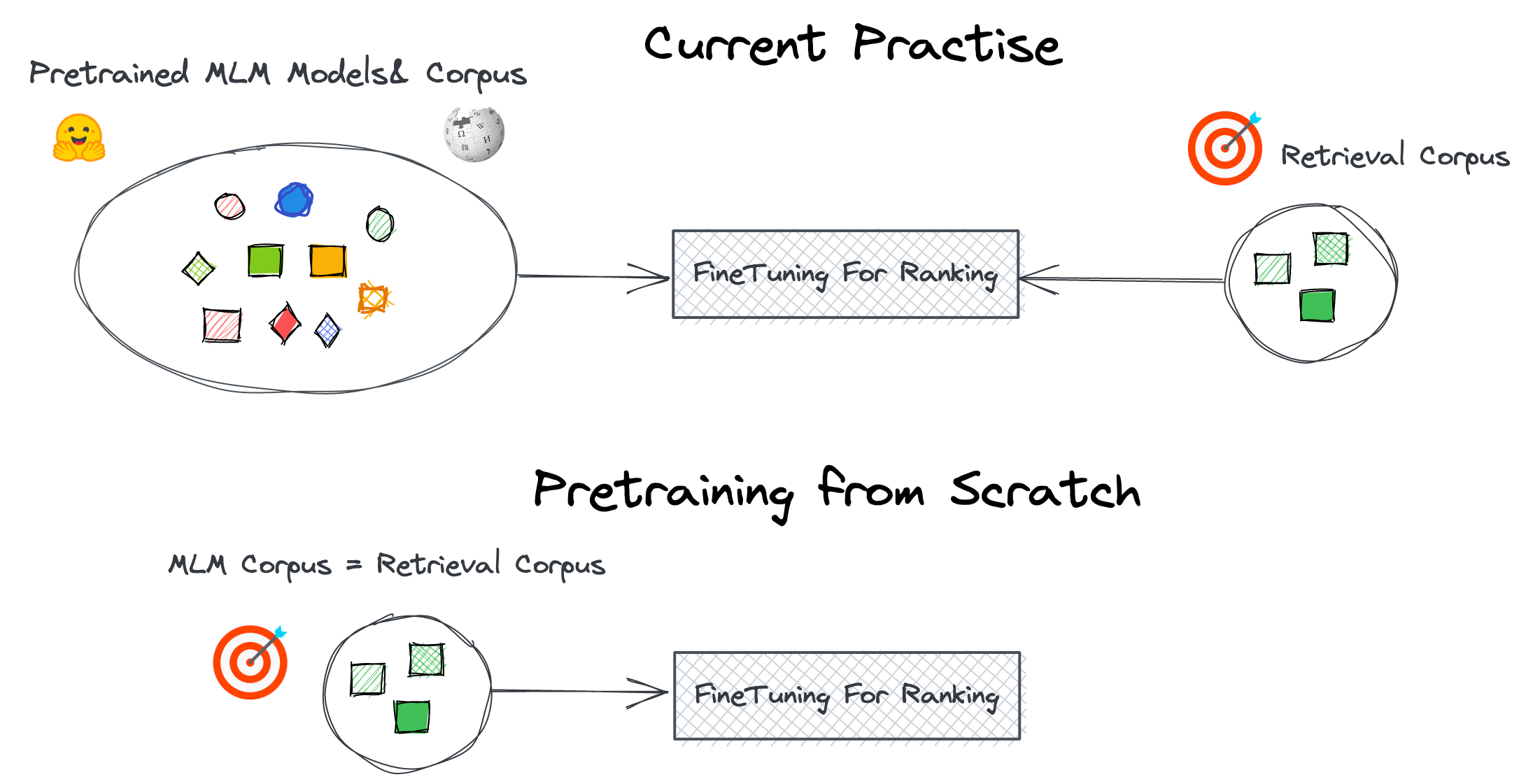}
    \caption{Pretraining from Scratch: MLM is only performed on the retrieval collection.}
    \label{fig:my_label}
\end{figure}

Therefore, by comparing these two approaches, we hope to better understand how large external pretraining contributes to the final effectiveness. In this paper, our research question deals with general-purpose vs specific-purpose model. The mainstream approach is to adopt the general purpose model, by simply adapting it to an IR task. On contrary, this paper investigates specific purpose models to assess their effectiveness (cf. Figure \ref{fig:my_label}). In a nutshell, would pretraining from scratch work for IR models?
More specifically, we look at the following research questions:
\begin{enumerate}
    \item Do we need an external pretrained language model for Information Retrieval?
    \item Do models pretrained on target collections generalize to other domains and tasks?
    \item Can we take advantage of pretraining for specialized domains and non-English languages?
    \item Does efficient pretraining allow us to use recent architectural advances of transformers?
\end{enumerate}

To answer these questions,  we compare the performance of standards models, such as BERT pretrained in Wikipedia and Book Corpus, to pretraining for scratch on the retrieval collections. For instance, we use the classical MSMARCO~\cite{nguyen2016ms} dataset. By pretraining only on MSMARCO, we observe if there is indeed a benefit when pretraining on Wikipedia. Furthermore, we assess whether such models still generalise well on the BEIR dataset \cite{beir}. Indeed, one advantage of pretraining on a large collection could be better generalization in zero shot settings.
For specialized domains, such as health and biomedical data, we use the TripClick~\cite{rekabsaz2021fairnessir} dataset and compare pretraining from scratch to the performance of SciBERT \cite{scibert} and PubMedBERT \cite{Gu_2022}. Similarly, by pretraining models solely on TripClick, we compare the performance against models trained on a much larger collection (i.e PubMed). Finally, we will extend the previous experiments on non-English datasets, on  Mr. TyDi~\cite{zhang2021mr} dataset by comparing to MContriever ~\cite{izacard2021contriever}, which relied on a pretraining on large dataset with many languages. By conducting those experiments, we can assess whether an external pretrained language model is needed for Information Retrieval and if pretraining from scratch is an interesting alternative.

We will focus our study on BERT~\cite{bert} and DistilBERT~\cite{sanh2019distilbert} architectures since they are the most popular ones in IR. Furthermore, we will measure the effectiveness of different architecture: \textit{dense} bi-encoders (\cite{sentence_bert,lin2020IRBertReview}), \textit{SPLADE}, a state of the art sparse model~\cite{splade_sigir22,efficiency_splade_22} and cross-encoders for reranking~\cite{passage_ranking}.

\begin{table*}[t!]
\scriptsize
\begin{center}

    \caption{Model Comparison.}
    \label{tab:models}
\begin{tabular}{c|c|c|cc}
 \textbf{ }& BERT & DistilBERT & MLM 6L & MLM 12L \\
\toprule
 MLM Collection & \tiny{Wikipedia, BookCorpus} & \tiny{Wikipedia, BookCorpus}    & \multicolumn{2}{c}{\tiny{MSMARCO or Mr. TyDi or TripClick}}       \\
 Pretraining Size (words)     & 3,300M         &  3,300M   &  \multicolumn{2}{c}{From 100M to 500M}              \\
 \# params                &  110M  & 66M      &  67M      &  110M            \\
 \# Layers                &    12 & 6         & 6     & 12            \\
 \#  number\_heads          &   12 & 16          &  16    & 12            \\
\end{tabular}
\end{center}
\end{table*}

\subsection{Pretraining}
In the following we always pretrain at least two types of models from scratch: a model with 12 layers based on the BERT architecture~\cite{bert}  and one with 6 layers based on DistilBERT~\cite{sanh2019distilbert}. We use BERT and DistilBERT as baselines all along the article. Table \ref{tab:models} summarises the main model characteristics we will consider.
For pretraining from scratch, we always fix the vocabulary size to 32k (slightly larger than BERT's 30.5k), using wordpiece~\cite{wu2016googles} to find the most common tokens of the target collection. 
We refer to those models as MLM 12L and MLM 6L. 

We also use 2 models built on a setting called, MLM+FLOPS 12L and MLM+FLOPS 6L, by adding the FLOPS regularization~\cite{paria2020minimizing}, which helps to (pre)condition PLM for usage with SPLADE as proposed in ~\cite{efficiency_splade_22}.
The standard MLM loss is modified as follows: the MLM logits go through the SPLADE activation function (i.e. $ \log(1+ReLU(y_\text{logits}))$), which  defines an MLM loss over a sparse set of logits. 
Finally, another term (FLOPS regularization) is added to force the logits to not only be nonnegative, but actually be sparse. As in SPLADE, a max pooling of the overall sentence is done to get a representation at the word level. On this final representation, the FLOPS regularization forces sparsification (and uniformity) over the overall vocabulary. The total loss is given by $\ell_\text{MLM} + \ell_\text{MLM-SPLADE} + \ell_\text{FLOPS}$.

Pretaining time is between 6 hours to 1 day depending on the models and collections on 8 NVIDIA A100 80 Gb, compared to the original computational cost of BERT (3 days using 64 TPUs), and the cost of finetuning (around 1 day on 4 V100 32Gb), we consider our pretraining cost to be reasonable.









\section{Experiments}
\input{experiments}

\section{Conclusion}
Foundation models come with the promise to be highly general and modular. It is believed that they contain a wide ``knowledge'' due to their pretraining on a large collection, which is then believed to be the source of  their improved performance. We have examined how this pretraining collection influence the performance of IR models. Our research question was to assess how much of this implicit knowledge, beneficial to the final performance, comes from pretraining on a large external collection. This is why we have experimented on a variety of collections, domains and languages to study how pretraining from scratch actually performed compared to their \textit{de facto} approach of simple finetuning. 

While we were expecting the standard pretrained models to work better, we surprisingly revealed that pretraining from scratch works better for first-stage retrieval on MSMARCO, TripClick and several non-English languages on the Mr. TyDi benchmark. In particular, the FLOPS regularization played a critical role in those results, suggesting that regularization or better pretraining techniques could further improve the results. Furthermore, pretrained models from scratch also behave well in the zero shot scenario for sparse models such as SPLADE. Nevertheless, pretraining from a large collection has a slight advantage when training rerankers. 

Overall, these results, specific to IR, challenge the foundation model hypothesis for small models, ie that a more general model encapsulating the world knowledge would be better than a smaller one in a specific domain application. Furthermore, our study makes a contribution to the debate between general purpose and specific purpose models.
In a way, our experiments showed that less is more. In addition, pretrained language models come also with many challenges such as the societal bias in the data they have been trained on.
We hope that our study could convince practitioners, both from industry and academia, to reconsider specific purpose models by pretraining from scratch. Last but not least, doing so enable to better control efficiency, data bias and replicability, which are key research questions for the IR community.

\bibliographystyle{splncs04}
\bibliography{ecir_pretraining}

\end{document}

%% file: experiments.tex
Our research question is to assess whether models fully trained on the target collection perform as well as the generic DistilBERT or BERT models. First, we check the results on the general collection MSMARCO~\cite{nguyen2016ms} (RQ1) and how the models trained on it generalize (RQ2) to a zero-shot scenario in BEIR~\cite{beir}. We then verify if the results generalize to more specialized collections and non-English languages (RQ3). Finally, we take advantage of the fact that we are training models from scratch to test variants of transformer architectures (RQ4). Additional finetuning details are available at the end of the respective sections.

\subsubsection{Experimental Setup}
\label{sec:marcopre}
Pretraining is performed using 8 NVIDIAs A100 80Gb either on MS-MARCO (RQ1 and RQ2), or TripClick (RQ3), or Mr. TyDi dataset (RQ3). We always use a learning rate of 1e-4. In the case of the MSMARCO collection, we pretrain using MLM and MLM+FLOPS using the entire passage collection (8.8M) combined with the training queries (\~800k) for a total of (9.6M ``documents''), while for TripClick we separate the documents into training and validation (90/10 split). For all collections, the batch size per GPU is either 150 (12L) or 200 (6L). We use an exponential warmup for FLOPS of 5k steps, a warmup of 1k steps for the logits and a learning rate warmup of 10k steps. The $\lambda$ factor of FLOPS is set to $1e-3$, the max length before truncation to 256 tokens and the networks are pretrained for 125k steps.
For Mr. TyDi, we pretrain 3 networks (Arabic, Russian, Japanese) using MLM+FLOPS on the entire language's passage collection (2M for Arabic, 7M for Japanese and 9.6M for Russian) with a batch size per GPU of 200. Finally for TripClick, the only difference is the number of epochs: 60 epochs, and the batch size per GPU is 256 for the 6L models, and  128 for the 12L models. 

For finetuning, we used 4 V100 32gb for a sparse model (SPLADE) and a dense neural bi-encoder model. Please note that \textbf{we do not use distillation} from a reranker as it would entail transferring information from an existing PLM. For SPLADE, we use the L1 regularization over queries (following~\cite{efficiency_splade_22})  with $\lambda_q=1e-3$, and for documents a FLOPS regularisation with $\lambda_d=1e-3$ ($\lambda_d=5e-4$ on TripClick) following \cite{efficiency_splade_22}. The learning rate is set up to  2$e{-5}$. 
In all of our tables, \textbf{superscripts} denote \textbf{significant differences} according to a paired Student's t-test with Bonferroni's correction and $p \le 0.05$ with the corresponding table row; MRR@10 and nDCG@10 have been multiplied by 100.


\subsection{RQ1: Are models fully trained on MSMARCO as good as models pretrained on a diverse collection set?}
\label{rq1}

The first question we want to address is whether models that are solely trained on MSMARCO are as good as models that have used external corpora for pretraining (i.e. BookCorpus and Wikipedia). We first investigate models trained as first stage rankers, either using a dense bi-encoder \cite{lin2020IRBertReview,karpukhin-etal-2020-dense} or sa SPLADE model.

We finetuned the models using negatives that were sampled from a previously trained SPLADE model. For each V100 we use a batch of the maximum size that does not exceed the total memory. Batches are constructed such that ``one element'' of the batch is composed of the query itself, a positive passage related to the query and 16 or 32 negatives, depending on the network size (the actual batch sizes per GPU varies from 54 to 102 depending on the pretrained model size). Finetuning is considered finished after two epochs (arbitrary decision based on initial experiments with a validation set).
Note that the finetuning setting for both first stage and cross-encoders are almost the same, except for the fact that cross-encoders do not use in-batch negatives and use a learning rate of 1e-4.

We report the retrieval flops (noted R-FLOPS) for SPLADE models, i.e., the number of floating point operations on the inverted index to return the list of documents for a given query.
The R-FLOPS metric is defined by an estimation of the average number of floating-point operations between a query and a document which is defined as the expectation $\mathbb{E}_{q,d} \left[ \sum_{j \in V} p_j^{(q)}p_j^{(d)} \right]$ where $p_j$ is the activation probability for token $j$ in a document $d$ or a query $q$. It is empirically estimated from a set of approximately $100$k development queries, on the MS MARCO collection. It is thus an indication of the inverted index sparsity and of the computational retrieval cost for a sparse retriever (which is different from the inference cost of the model).

First stage retrievers results are described in Table~\ref{tab:msmarco_sparse} for sparse models and Table~\ref{tab:msmarco_dense} for dense models. Surprisingly, models trained solely on MSMARCO with MLM+FLOPS actually \textbf{can perform statistically significantly better} than their counterparts pretrained over larger collections, in both sparse and dense scenarios, while there's no statistically significant difference when considering just MLM on MSMARCO vs MLM on larger corpora. This shows that not only pretraining does not seem to care for a diverse collection, but that by focusing only on ``off-the-shelf models'' we could be losing possible performance gains of better initialized models. Also note that less computing was used to pretrain MLM+FLOPS 6L compared to its DistilBERT counterpart. 

\input{tables/msmarco_splade}

\input{tables/msmarco_dense}

Finally, we also test in the case of reranking using cross-encoders. Results are available in Table~\ref{tab:rerankers}. Overall there's no statistical significant gain on the models pretrained with external collections.

\input{tables/msmarco_reranking}


\subsection{RQ2: Do models pretrained in MSMARCO generalize well on other collections?}

In the previous section, we considered only in-domain results. While they have shown that we can outperform (or at least keep comparable effectiveness) using solely the target collection, they could be masking a possible gap in out-of-domain data. In order to verify that models solely pretrained and fine-tuned on MSMARCO do not lose effectiveness in out-of-domain data, we report results under the zero-shot BEIR benchmark in Table~\ref{tab:beir}. We actually observe a small boost in effectiveness on sparse retrieval when using models solely trained on MSMARCO, while on Dense there's a more apparent decrease of performance. The biggest difference for dense models is on the TREC Covid dataset which is far from the MSMARCO collection. Given the nature of the BEIR benchmark (mean over 13 datasets), those differences may not be significative.\footnote{We could not find in the literature an easy/practical way to perform statistical significance testing over BEIR.}

\input{tables/beir}

\subsection{RQ3: Can we take advantage of that pretraining from scratch in collections of specialized domains/languages}

\subsubsection{Domain Specific IR on TripClick}
The TripClick collection \cite{rekabsaz2021fairnessir} contains approximately 1.5 millions MEDLINE documents (title and abstract), and 692,000 queries. The test set is divided into three categories of queries: Head, Torso and Tail (decreasing frequency), which contain 1,175 queries each. For the Head queries a DCTR click model~\cite{craswell2008experimental} was employed to created relevance signals, the other two sets use raw clicks. \cite{hofstaetter2022tripclick} showed that the original triplets were too noisy, and released a new training set which we use in this experiment (10 millions triplets). 



As pretrained models, we use the off-the-shelf BERT, DistilBERT models, and similarly to \cite{hofstaetter2022tripclick} SciBERT \cite{scibert} and PubMedBERT \cite{Gu_2022} which are both using a similar architecture to Bert (12 layers) and were pretrained using scientific documents (from where they extract their vocabulary). We consider them as off-the-shelf domain-specific models.
While for finetuning, we use a batch size of 200 queries and only one negative per query, taking advantage solely of in-batch negatives for 90,000 iterations, which is equivalent to 1.8 epochs.

As Table~\ref{tab:tripclick} shows, models pretrained from scratch compete well against generic off-the-shelf models as well as against specialized  ones.
For this collection, the dense models perform better than the sparse ones. The conclusions depend on the model type: sparse or dense. For the sparse models, we see that at least one model based on pre-training from scratch outperforms  off-the-shelf models such as BERT and  DistilBERT, as well as both off-the-shelf domain-specific models (Scibert, PubMedBert). Regarding the dense architecture, pretrained models from scratch are on par with  off-the-shelf domain-specific models which required much more training data (1.4B tokens for Scibert against 300M for TripClick). 
Pre-training from scratch allows for selecting the most suitable language model according to the fine-tuning architecture (sparse or dense). In our case a 6 layer language model is far better for SPLADE while a 12 layer better fits a dense model.
We added at the bottom of Table~\ref{tab:tripclick} the results from \cite{hofstaetter2022tripclick}, which show that our implementation choices are very competitive\footnote{We were not able to find the parameters used in the experiments.} for the sparse as well as for the dense models. 

\input{tables/tripclick}

\subsubsection{Mr. TyDi}
is a multilingual dataset for monolingual retrieval composed of 11 typologically diverse languages~\cite{zhang2021mr}. For this study we focus on the three non-English languages with the most training data on Mr. TyDi: i) Arabic; ii) Russian; iii) Japanese. Note that, there is a large amount of data available for these languages (even outside of Mr. TyDi), but PLMs are not as well studied as in English. The main consensus seems to be that in this case one should focus on multi-lingual data, for which most is based on english as an anchor, as it is the case for many previous works~\cite{izacard2021contriever,nair2022learning,nair2022transfer,zhang2021mr,conneau2019unsupervised}. We challenge this notion, by: a) using monolingual models; b) pretraining solely on Mr. TyDi.

We follow the fine-tuning protocol of MContriever ~\cite{izacard2021contriever}, where they first finetune the model for retrieval on MMarco~\cite{bonifacio2021mmarco}, a translated version of MSMARCO in multiple languages, for which we only use the target language for a given model (Arabic, Russian or Japanese) and then finally fine-tune on the Mr. TyDi collection. In our case we perform a three-step training, first step on MMarco using negatives sampled with the pretrained model (or MContriever for the baseline). We then perform two steps of training on Mr. TyDi, first using negatives extracted first from the MMarco finetuned model and second from the first stage of Mr.TyDi finetuning. Batch composition follows the previous MSMARCO finetuning, with a batch size of 3 queries 32 negatives (the actual batch sizes per GPU is thus 102). Finetuning stopped after two epochs.


Results are available in Table~\ref{tab:mrtydi}. Compared to the previous state of the art~\cite{izacard2021contriever}\footnote{Note that MContriever TyDi (first row) is not available, statistical tests cannot be performed. We do our best to evaluate fairly under our training setting (second row)}, which is a dense retriever pretrained in a specific fashion on a much larger collection\footnote{We suspect they use more compute, but could not find accurate compute information.}, we show statistically significant improvements on all languages while using solely the Mr. TyDi and MMarco collections on the target language. However, it is important to note that differently from~\cite{izacard2021contriever} we did not actually test  yet on all languages and thus can only evaluate the pretraining effect on these three languages, which are the three largest from Mr. TyDi. 

\input{tables/mrtidy}




\subsection{RQ4: Impact of Architectures}
Finally, one advantage of pretraining models from scratch is the fact that we can more easily experiment with different architectures. Indeed, considering that most IR works use a variant of BERT (either RoBERTA, DistilBERT or BERT) it raises a question whether variants of transformer architectures, benchmarked in NLP, could actually improve IR. To address this question, we use the sparse retrieval setting from RQ1, but this time also consider using the DeBERTa architecture~\cite{he2021deberta} which beats BERT on many NLP tasks. Results are presented in Tables~\ref{tab:msmarco_sparse_deb} and~\ref{tab:tripclickdeberta}. Much to our dismay, we did not actually see major improvements using these architectural changes, we thus leave as future work how to better include these changes within first stage rankers.

\input{tables/msmarco_deberta}
\input{tables/tripclick_deberta}

%% file: tables/msmarco_splade.tex
\begin{table*}[t!]
\centering
    \caption{Comparison on MSMARCO of  first stage sparse neural (SPLADE) models. $^\dagger$ indicates a method pretrained solely on MSMARCO.}
    \label{tab:msmarco_sparse}
    \resizebox{0.95\textwidth}{!}{
        \begin{tabular}{r | l | c | c c | c c | c c}
        \toprule
        \multirow{2}{*}{\textbf{\#}} & \multirow{2}{*}{Pretrained Model} & \multirow{2}{*}{R-FLOPS}  & \multicolumn{2}{c|}{MSMARCO dev} & \multicolumn{2}{c|}{TREC DL 19} & \multicolumn{2}{c}{TREC DL 20} \\
        &  & & MRR@10 & R@1k & nDCG@10 & R@1k & nDCG@10 & R@1k \\
        \midrule
a &Distilbert & 1.10 &0.373\hphantom{$^{bcde}$} &0.975$^{b}$\hphantom{$^{cde}$} &\textbf{0.732}\hphantom{$^{bcde}$} &\textbf{0.853}\hphantom{$^{bcde}$} &0.708\hphantom{$^{bcde}$} &0.867$^{b}$\hphantom{$^{cde}$} \\
b &BERT & 1.32 &0.367\hphantom{$^{acde}$} &0.968\hphantom{$^{acde}$} &0.727\hphantom{$^{acde}$} &0.832\hphantom{$^{acde}$} &0.699\hphantom{$^{acde}$} &0.842\hphantom{$^{acde}$} \\
c &MLM 6L$^\dagger$ & 8.2 &0.370\hphantom{$^{abde}$} &\textbf{0.982}$^{ab}$\hphantom{$^{de}$} &0.701\hphantom{$^{abde}$} &0.847\hphantom{$^{abde}$} &0.700\hphantom{$^{abde}$} &\textbf{0.877}\hphantom{$^{abde}$} \\
d &MLM+Flops 6L$^\dagger$ & 0.72 &\textbf{0.382}$^{abc}$\hphantom{$^{e}$} &0.979$^{b}$\hphantom{$^{ace}$} &0.698\hphantom{$^{abce}$} &0.836\hphantom{$^{abce}$} &0.701\hphantom{$^{abce}$} &0.872\hphantom{$^{abce}$} \\
e &MLM+Flops 12L$^\dagger$ & 0.97 &0.379$^{bc}$\hphantom{$^{ad}$} &0.980$^{ab}$\hphantom{$^{cd}$} &0.709\hphantom{$^{abcd}$} &0.835\hphantom{$^{abcd}$} &\textbf{0.709}\hphantom{$^{abcd}$} &0.865\hphantom{$^{abcd}$} \\
        \bottomrule

    \end{tabular}   
    }
\end{table*}

%% file: tables/msmarco_dense.tex
\begin{table*}[t!]
\centering
    \caption{Comparison on MSMARCO of  first stage dense neural (DPR) models. $^\dagger$ indicates a method pretrained solely on MSMARCO.}
    \label{tab:msmarco_dense}
    \resizebox{0.95\textwidth}{!}{
        \begin{tabular}{r | c | c c | c c | c c}
        \toprule
        \multirow{2}{*}{\textbf{\#}} & \multirow{2}{*}{Pretrained Model} & \multicolumn{2}{c|}{MSMARCO dev} & \multicolumn{2}{c|}{TREC DL 19} & \multicolumn{2}{c}{TREC DL 20} \\
        &  & MRR@10 & R@1k & nDCG@10 & R@1k & nDCG@10 & R@1k \\
        \midrule

a &Distilbert &0.342\hphantom{$^{bcde}$} &0.961\hphantom{$^{bcde}$} &0.673\hphantom{$^{bcde}$} &0.774\hphantom{$^{bcde}$} &0.670\hphantom{$^{bcde}$} &0.816\hphantom{$^{bcde}$} \\
b &BERT &0.347\hphantom{$^{acde}$} &0.961\hphantom{$^{acde}$} &\textbf{0.697}\hphantom{$^{acde}$} &0.785\hphantom{$^{acde}$} &\textbf{0.682}\hphantom{$^{acde}$} &0.809\hphantom{$^{acde}$} \\
c &MLM 6L$^\dagger$ &0.346\hphantom{$^{abde}$} &0.968$^{ab}$\hphantom{$^{de}$} &0.664\hphantom{$^{abde}$} &0.783\hphantom{$^{abde}$} &0.657\hphantom{$^{abde}$} &0.818\hphantom{$^{abde}$} \\
d &MLM+FLOPS 6L $^\dagger$ &0.349\hphantom{$^{abce}$} &0.968$^{ab}$\hphantom{$^{ce}$} &0.670\hphantom{$^{abce}$} &0.781\hphantom{$^{abce}$} &0.668\hphantom{$^{abce}$} &0.837\hphantom{$^{abce}$} \\
e &MLM+FLOPS 12L$^\dagger$ &\textbf{0.352}$^{a}$\hphantom{$^{bcd}$} &\textbf{0.969}$^{ab}$\hphantom{$^{cd}$} &0.672\hphantom{$^{abcd}$} &\textbf{0.800}\hphantom{$^{abcd}$} &0.680\hphantom{$^{abcd}$} &\textbf{0.848}$^{bc}$\hphantom{$^{ad}$} \\
        \bottomrule

    \end{tabular}   
    }
\end{table*}

%% file: tables/msmarco_reranking.tex
\begin{table*}[t!]
\begin{center}

    \caption{Comparison of rerankers on MSMARCO. Models with $\dagger$ were pretrained solely on MSMARCO.}
    \label{tab:rerankers}
\resizebox{0.95\textwidth}{!}{
\begin{tabular}{c|c|c|c|c}
\toprule
\multirow{2}{*}{\textbf{\#}} & \multirow{2}{*}{Pretrained Model}& MSMARCO dev& TREC DL 2019 & TREC DL 2020 \\ 
& & \textbf{MRR@10}& \textbf{nDCG@10}& \textbf{nDCG@10} \\ 
\midrule
a &Without reranking (First stage)  &0.384\hphantom{$^{bcdef}$} &0.718\hphantom{$^{bcdef}$} &\textbf{0.737}\hphantom{$^{bcdef}$} \\
b &Distilbert &0.396$^{af}$\hphantom{$^{cde}$} &\textbf{0.764}\hphantom{$^{acdef}$} &0.734\hphantom{$^{acdef}$} \\
c &Bert &\textbf{0.404}$^{abf}$\hphantom{$^{de}$} &0.750\hphantom{$^{abdef}$} &0.737\hphantom{$^{abdef}$} \\
d &MLM 6L $^\dagger$ &0.398$^{af}$\hphantom{$^{bce}$} &0.743\hphantom{$^{abcef}$} &0.716\hphantom{$^{abcef}$} \\
e &MLM+Flops 6L $^\dagger$ &0.396$^{af}$\hphantom{$^{bcd}$} &0.724\hphantom{$^{abcdf}$} &0.736\hphantom{$^{abcdf}$} \\
f &MLM+Flops 12L $^\dagger$ &0.381\hphantom{$^{abcde}$} &0.730\hphantom{$^{abcde}$} &0.722\hphantom{$^{abcde}$} \\
\bottomrule
\end{tabular}
}
\end{center}
\end{table*}

%% file: tables/beir.tex
\begin{table*}[t!]
\centering
    \caption{Experiments on zero-shot retrieval on BEIR (nDCG@10) with models fine-tuned on MSMARCO. Models with $\dagger$ were pretrained solely on MSMARCO.}
    \label{tab:beir}
    \resizebox{0.95\textwidth}{!}{
    
\begin{tabular}{lccccc|ccccc}
\toprule
{} &  \multicolumn{5}{c|}{SPLADE} &  \multicolumn{5}{c}{Dense} \\
{} &  Distilbert &  Bert& M 6L$^\dagger$ &  M+F 6L$^\dagger$ &  M+F 12L$^\dagger$ &  Distilbert &  Bert & M 6L$^\dagger$ &  M+F 6L$^\dagger$ &  M+F 12L$^\dagger$ \\
\midrule
arguana          &                                      45.30 &                                44.30 &                                        \textbf{46.90} &                               42.90 &                                     45.40 &                                          34.00 &                                      37.30 &                                          40.20 &                                          37.40 &                                          40.70 \\
climate-fever    &                                      14.90 &                                15.30 &                                        14.50 &                               15.70 &                                     15.40 &                                          16.50 &                                      15.90 &                                          15.40 &                                          15.30 &                                          \textbf{17.50} \\
dbpedia-entity   &                                      \textbf{39.10} &                                38.80 &                                        38.30 &                               38.80 &                                     37.90 &                                          31.50 &                                      31.70 &                                          31.10 &                                          31.70 &                                          30.60 \\
fever            &                                      \textbf{73.40} &                                71.20 &                                        68.60 &                               72.30 &                                     70.70 &                                          70.80 &                                      70.20 &                                          59.10 &                                          59.90 &                                          56.10 \\
fiqa             &                                      31.20 &                                29.90 &                                        \textbf{33.20} &                               32.30 &                                     31.70 &                                          25.70 &                                      24.10 &                                          27.30 &                                          27.00 &                                          27.60 \\
hotpotqa         &                                      66.90 &                                65.50 &                                        64.00 &                               67.30 &                                     \textbf{67.60} &                                          49.70 &                                      50.20 &                                          47.30 &                                          46.40 &                                          47.60 \\
nfcorpus         &                                      33.40 &                                31.30 &                                        35.40 &                               \textbf{35.90} &                                     34.70 &                                          26.40 &                                      25.80 &                                          28.80 &                                          28.70 &                                          28.80 \\
nq               &                                      \textbf{51.40} &                                50.60 &                                        48.50 &                               49.20 &                                     49.40 &                                          46.00 &                                      47.10 &                                          41.90 &                                          41.50 &                                          42.90 \\
quora            &                                      77.20 &                                76.60 &                                        80.40 &                               78.10 &                                     73.80 &                                          78.50 &                                      82.20 &                                          82.20 &                                          \textbf{83.30} &                                          80.30 \\
scidocs          &                                      14.90 &                                14.90 &                                        14.40 &                               \textbf{15.10} &                                     14.50 &                                          11.40 &                                      11.80 &                                          10.70 &                                          11.20 &                                          11.00 \\
scifact          &                                      66.00 &                                64.70 &                                        \textbf{69.20} &                               69.20 &                                     69.00 &                                          52.00 &                                      54.80 &                                          56.70 &                                          57.30 &                                          56.50 \\
trec-covid       &                                      67.60 &                                \textbf{69.10} &                                        65.60 &                               64.70 &                                     68.00 &                                          66.10 &                                      65.70 &                                          56.00 &                                          48.90 &                                          49.90 \\
webis-touche2020 &                                      27.60 &                                27.00 &                                        24.70 &                               \textbf{28.50} &                                     26.20 &                                          22.20 &                                      23.50 &                                          23.20 &                                          19.70 &                                          19.30 \\ \midrule
mean             &                                      \textbf{46.84} &                                46.09 &                                        46.43 &                               \textbf{46.92} &                                     46.48 &                                          40.83 &                                      41.56 &                                          39.99 &                                          39.10 &                                          39.14 \\
\bottomrule
\end{tabular}

}
\end{table*}

%% file: tables/tripclick.tex
\begin{table*}[t!]
\centering
    \caption{Experiment on Tripclick with sparse (SPLADE) and dense models (nDCG@10).$^\dagger$ indicates a method pretrained solely on Tripclik.}
    \label{tab:tripclick}
    \resizebox{1.0\textwidth}{!}{
        \begin{tabular}{r | l | c  c   c   c  |c  c   c   c}
        \toprule   
       \multirow{2}{*}{\#}& \multirow{2}{*}{Pretrained Model}   &     \multicolumn{4}{c|}{SPLADE} &  \multicolumn{4}{c}{Dense}   \\ 
        &   &Head$_{dctr}$ &Head  &Torso & Tail    &Head$_{dctr}$ &Head  &Torso & Tail  \\
        \midrule
a &
Distilbert &
22.1$^{b}$\hphantom{$^{cefgh}$} &
30.3$^{bd}$\hphantom{$^{cefgh}$} &
23.9$^{b}$\hphantom{$^{cdefgh}$} &
24.8$^{bd}$\hphantom{$^{cefgh}$} &
24.9\hphantom{$^{bcdefgh}$} &
34.8\hphantom{$^{bcdefgh}$} &
29.2\hphantom{$^{bcdefgh}$} &
27.5\hphantom{$^{bcdefgh}$} \\
b &
BERT &
10.6\hphantom{$^{acdefgh}$} &
14.4\hphantom{$^{acdefgh}$} &
9.6\hphantom{$^{acdefgh}$} &
7.8\hphantom{$^{acdefgh}$} &
25.3\hphantom{$^{acdefgh}$} &
35.3\hphantom{$^{acdefgh}$} &
29.4\hphantom{$^{acdefgh}$} &
28.8$^{a}$\hphantom{$^{cdefgh}$} \\
c &
PubMedBert &
22.5$^{bd}$\hphantom{$^{aefgh}$} &
31.0$^{bd}$\hphantom{$^{aefgh}$} &
24.5$^{b}$\hphantom{$^{aefgh}$} &
24.3$^{bd}$\hphantom{$^{aefgh}$} &
27.1$^{abef}$\hphantom{$^{dgh}$} &
37.6$^{abef}$\hphantom{$^{dgh}$} &
29.9$^{e}$\hphantom{$^{abdfgh}$} &
\textbf{30.8}$^{a}$\hphantom{$^{bcdegh}$} \\
d &
Scibert &
21.1$^{b}$\hphantom{$^{acefgh}$} &
28.4$^{b}$\hphantom{$^{acefgh}$} &
23.0$^{b}$\hphantom{$^{acefgh}$} &
22.2$^{b}$\hphantom{$^{acefgh}$} &
27.8$^{abef}$\hphantom{$^{cgh}$} &
38.1$^{abef}$\hphantom{$^{cgh}$} &
29.2\hphantom{$^{abcefgh}$} &
30.3$^{a}$\hphantom{$^{bcefgh}$} \\
e &
 MLM+F 6L$^\dagger$ &
\textbf{26.9}$^{abcdfgh}$\hphantom{} &
\textbf{36.7}$^{abcdfgh}$\hphantom{} &
\textbf{27.7}$^{abcdfgh}$\hphantom{} &
\textbf{27.2}$^{bcdfh}$\hphantom{$^{g}$} &
25.7\hphantom{$^{abcdfgh}$} &
35.7\hphantom{$^{abcdfgh}$} &
28.3\hphantom{$^{abcdfgh}$} &
29.7\hphantom{$^{abcdfgh}$} \\
f &
MLM 6L$^\dagger$ &
23.1$^{bd}$\hphantom{$^{acegh}$} &
31.9$^{bd}$\hphantom{$^{acegh}$} &
24.7$^{b}$\hphantom{$^{acdegh}$} &
23.4$^{b}$\hphantom{$^{acdegh}$} &
25.8\hphantom{$^{abcdegh}$} &
36.0\hphantom{$^{abcdegh}$} &
28.9\hphantom{$^{abcdegh}$} &
29.3\hphantom{$^{abcdegh}$} \\
g &
 MLM+F 12L$^\dagger$ &
23.7$^{abd}$\hphantom{$^{cefh}$} &
32.5$^{abd}$\hphantom{$^{cefh}$} &
26.2$^{abd}$\hphantom{$^{cefh}$} &
26.6$^{bdf}$\hphantom{$^{aceh}$} &
26.7$^{ab}$\hphantom{$^{cdedh}$} &
37.5$^{abef}$\hphantom{$^{cdh}$} &
29.9$^{e}$\hphantom{$^{abcdh}$} &
30.1$^{a}$\hphantom{$^{bcdefh}$} \\
h &
MLM 12L$^\dagger$ &
24.0$^{abcd}$\hphantom{$^{efg}$} &
32.6$^{abcd}$\hphantom{$^{efg}$} &
24.8$^{bd}$\hphantom{$^{acefg}$} &
24.6$^{bdf}$\hphantom{$^{aceg}$} &
\textbf{28.0}$^{abefg}$\hphantom{$^{cd}$} &
\textbf{38.6}$^{abef}$\hphantom{$^{dcg}$} &
\textbf{30.2}$^{e}$\hphantom{$^{abcdg}$} &
30.4$^{a}$\hphantom{$^{bcdefg}$} \\

        \bottomrule
        &Scibert\cite{hofstaetter2022tripclick}   & -&- &- &- &24.3\hphantom{$^{abcdefgh}$} &- 	&	- &-\\ 
        &PubMebBert \cite{hofstaetter2022tripclick} &- & -&- &-&	23.5\hphantom{$^{abcdefgh}$}&	-& -&- \\
        &BM25 \cite{hofstaetter2022tripclick}	&- & -&- &-&  14.0\hphantom{$^{abcdefgh}$}	 & -	&- & -\\
       \hline
    \end{tabular}   
    }
\end{table*}

%% file: tables/mrtidy.tex
\begin{table*}
\centering
\caption{Comparison, on the Mr.TyDi dataset, of models trained from scratch against models pretrained in a large external collection with Contriever.$^\dagger$ means a method pretrained solely on MrTiDy.}
\label{tab:mrtydi}
\resizebox{0.95\textwidth}{!}{
\begin{tabular}{c|c|c|c|c|c|c|c}
\toprule
\textbf{\#}
& \textbf{Pretrained Model}& \multicolumn{2}{c|}{Arabic} & \multicolumn{2}{c|}{Russian} & \multicolumn{2}{c}{Japanese} \\ 
& & \textbf{MRR@100}& \textbf{R@100}& \textbf{MRR@100}& \textbf{R@100}& \textbf{MRR@100}& \textbf{R@100} \\ 
\midrule
 & MContriever~\cite{izacard2021contriever} & 72.4 \hphantom{$^{bc}$} & 94.0 \hphantom{$^{bc}$} & 59.7 \hphantom{$^{bc}$} & 92.4 \hphantom{$^{bc}$} & 54.9 \hphantom{$^{bc}$} & 88.8 \hphantom{$^{bc}$} \\
\midrule
a &MContriever (reproduced) &72.7\hphantom{$^{bc}$} &93.6\hphantom{$^{bc}$} &59.9$^{b}$\hphantom{$^{c}$} &91.6\hphantom{$^{bc}$} &49.9$^{b}$\hphantom{$^{c}$} &85.2$^{b}$\hphantom{$^{c}$} \\
b &MLM+FLOPS 6L DPR$^\dagger$ &73.4\hphantom{$^{ac}$} &\textbf{94.9}\hphantom{$^{ac}$} &56.2\hphantom{$^{ac}$} &91.6\hphantom{$^{ac}$} &32.9\hphantom{$^{ac}$} &62.0\hphantom{$^{ac}$} \\
c &MLM+FLOPS 6L SPLADE$^\dagger$ &\textbf{75.7}$^{ab}$\hphantom{} &94.2\hphantom{$^{ab}$} &\textbf{65.0}$^{ab}$\hphantom{} &\textbf{93.4}$^{ab}$\hphantom{} &\textbf{56.3}$^{ab}$\hphantom{} &\textbf{88.9}$^{ab}$\hphantom{} \\
\bottomrule
\end{tabular}
}
\label{tab:results}
\end{table*}

%% file: tables/msmarco_deberta.tex
\begin{table*}[t!]
\centering
    \caption{Comparison on MSMARCO of  first stage sparse neural (SPLADE) models using different architectures. All methods are pretrained solely on MSMARCO.}
    \label{tab:msmarco_sparse_deb}
    \resizebox{0.95\textwidth}{!}{
        \begin{tabular}{r | l | c | c c | c c | c c}
        \toprule
        \multirow{2}{*}{\textbf{\#}} & \multirow{2}{*}{Pretrained Model} & \multirow{2}{*}{R-FLOPS}  & \multicolumn{2}{c|}{MSMARCO dev} & \multicolumn{2}{c|}{TREC DL 19} & \multicolumn{2}{c}{TREC DL 20} \\
        &  & & MRR@10 & R@1k & nDCG@10 & R@1k & nDCG@10 & R@1k \\
        \midrule

a &BERT MLM+FLOPS 6L & 0.72 &\textbf{0.382}$^{d}$\hphantom{$^{bc}$} &0.979\hphantom{$^{bcd}$} &0.698\hphantom{$^{bcd}$} &0.836\hphantom{$^{bcd}$} &0.701\hphantom{$^{bcd}$} &\textbf{0.872}\hphantom{$^{bcd}$} \\
b &BERT MLM+FLOPS 12L& 0.97 &0.379\hphantom{$^{acd}$} &\textbf{0.980}\hphantom{$^{acd}$} &0.709\hphantom{$^{acd}$} &0.835\hphantom{$^{acd}$} &0.709\hphantom{$^{acd}$} &0.865\hphantom{$^{acd}$} \\
\midrule
c &DeBERTa MLM+FLOPS 6L & 0.81 &0.376\hphantom{$^{abd}$} &0.979\hphantom{$^{abd}$} &0.700\hphantom{$^{abd}$} &\textbf{0.845}\hphantom{$^{abd}$} &0.701\hphantom{$^{abd}$} &0.863\hphantom{$^{abd}$} \\
d &DeBERTa MLM+FLOPS 12L& 0.79 &0.373\hphantom{$^{abc}$} &0.978\hphantom{$^{abc}$} &\textbf{0.716}\hphantom{$^{abc}$} &0.839\hphantom{$^{abc}$} &\textbf{0.735}\hphantom{$^{abc}$} &0.871\hphantom{$^{abc}$} \\

        \bottomrule

    \end{tabular}   
    }
\end{table*}

%% file: tables/tripclick_deberta.tex
\begin{table*}[t!]
\centering
    \caption{Comparison on TripClick of first stage sparse (SPLADE) and dense models using different architectures. All methods are pretrained solely on TripClick.}
    \label{tab:tripclickdeberta}
    \resizebox{0.99\textwidth}{!}{
        \begin{tabular}{r | l | c | c  | c  | c  |c | c  | c  | c}
        \toprule   
       \multirow{2}{*}{\#}& \multirow{2}{*}{Pretrained Model}   &     \multicolumn{4}{c|}{SPLADE} &  \multicolumn{4}{c}{Dense}   \\ 
        &   & Head$_{dctr}$ &Head  &Torso & Tail    &Head$_{dctr}$ &Head  &Torso & Tail  \\
       \hline
        \midrule

a &
BERT MLM+FLOPS 6L &
26.9$^{bcdefh}$\hphantom{$^{g}$} &
36.7$^{bcdfh}$\hphantom{$^{aeg}$} &
27.7$^{bdfh}$\hphantom{$^{aeg}$} &
27.2$^{bdfh}$\hphantom{$^{aceg}$} &
25.7$^{e}$\hphantom{$^{abcdfgh}$} &
35.7$^{e}$\hphantom{$^{abcdfgh}$} &
28.3\hphantom{$^{abcdefgh}$} &
29.7$^{e}$\hphantom{$^{abcdfgh}$} \\
b &
BERT MLM 6L &
23.1$^{}$\hphantom{$^{acdefgh}$} &
31.9$^{}$\hphantom{$^{acdefgh}$} &
24.7$^{}$\hphantom{$^{acdefgh}$} &
23.4\hphantom{$^{acdefgh}$} &
25.8$^{e}$\hphantom{$^{acdfgh}$} &
36.0$^{e}$\hphantom{$^{acdfgh}$} &
28.9$^{}$\hphantom{$^{abcdefgh}$} &
29.3$^{}$\hphantom{$^{abcdefgh}$} \\
c &
BERT MLM+FLOPS 12L &
23.7$^{f}$\hphantom{$^{abdegh}$} &
32.5$^{f}$\hphantom{$^{abdegh}$} &
26.2$^{fh}$\hphantom{$^{abdeg}$} &
26.6$^{bfh}$\hphantom{$^{acdeg}$} &
26.7$^{ef}$\hphantom{$^{abcdgh}$} &
37.5$^{abef}$\hphantom{$^{cdgh}$} &
29.9$^{aef}$\hphantom{$^{bcdgh}$} &
30.1$^{e}$\hphantom{$^{abcdfgh}$} \\
d &
BERT MLM 12L &
24.0$^{fh}$\hphantom{$^{abceg}$} &
32.6$^{f}$\hphantom{$^{abcegh}$} &
24.8$^{}$\hphantom{$^{abcefgh}$} &
24.6$^{}$\hphantom{$^{abcefgh}$} &
\textbf{28.0}$^{abcefgh}$\hphantom{} &
\textbf{38.6}$^{abefgh}$\hphantom{c} &
\textbf{30.2}$^{aeh}$\hphantom{$^{bcfg}$} &
\textbf{30.4}$^{e}$\hphantom{$^{abcfgh}$} \\
\hline
e &
DeBERTa MLM FLOPS 6L &
26.1$^{bcdfh}$\hphantom{$^{ag}$} &
35.9$^{bcdfh}$\hphantom{$^{ag}$} &
29.0$^{bcdfh}$\hphantom{$^{g}$} &
28.4$^{bcdfh}$\hphantom{$^{ag}$} &
24.3\hphantom{$^{abcdfgh}$} &
34.0\hphantom{$^{abcdfgh}$} &
27.8\hphantom{$^{abcdfgh}$} &
27.5\hphantom{$^{abcdfgh}$} \\
f &
DeBERTa MLM 6L  &
22.2\hphantom{$^{abcdegh}$} &
30.7\hphantom{$^{abcdegh}$} &
23.3\hphantom{$^{abcdegh}$} &
23.3\hphantom{$^{abcdegh}$} &
24.9\hphantom{$^{abcdegh}$} &
34.7\hphantom{$^{abcdegh}$} &
28.7\hphantom{$^{abcdegh}$} &
28.9\hphantom{$^{abcdegh}$} \\
g &
DeBERTa MLM FLOPS 12L &
\textbf{27.0}$^{bcdfh}$\hphantom{$^{a}$} &
\textbf{37.5}$^{bcdefh}$\hphantom{$^{a}$} &
\textbf{29.4}$^{abcdfh}$\hphantom{$^{e}$} &
\textbf{28.7}$^{abcdfh}$\hphantom{$^{e}$} &
26.6$^{ef}$\hphantom{$^{abcdh}$} &
36.7$^{ef}$\hphantom{$^{abcdh}$} &
29.2\hphantom{$^{abcdefh}$} &
29.5\hphantom{$^{abcdefh}$} \\
h &
DeBERTa MLM 12L &
22.6\hphantom{$^{abcdefg}$} &
31.3\hphantom{$^{abcdefg}$} &
23.8\hphantom{$^{abcdefg}$} &
23.3\hphantom{$^{abcdefg}$} &
26.4$^{ef}$\hphantom{$^{abcdg}$} &
36.4$^{ef}$\hphantom{$^{abcdg}$} &
28.6\hphantom{$^{abcdefg}$} &
29.0\hphantom{$^{abcdefg}$} \\
\bottomrule

\bottomrule
    \end{tabular}   
    }
\end{table*}